\documentclass[12pt]{article}
\usepackage[a4paper,margin=1in]{geometry}
\usepackage{amsmath, amssymb}
\usepackage{float}
\usepackage{tikz}
\usepackage{feynmp-auto}
\allowdisplaybreaks
\usepackage{tikz}
\usepackage{tikz-feynman}
\tikzfeynmanset{compat=1.1.0}
\usepackage{tikz-feynman}
\usetikzlibrary{positioning}

\begin{document}

\title{Radiative $\mu-\tau$ Corrections and Renormalization of Neutrino Mass Operators in Type II Seesaw Models}
\author{Gayatri Ghosh \\ \small Department of Physics, Cachar College \\ \small gayatrighsh@gmail.com}

\maketitle

\begin{abstract}
We explore the impact of radiative $\mu-\tau$ corrections on the renormalization of neutrino mass operators in the Type II Seesaw framework, incorporating both dimension-five and dimension-six operators. Using renormalization group equations (RGE), we analyze the evolution of flavor coupling matrices and their deviations from $\mu-\tau$ symmetric configurations due to quantum corrections. Given the stringent constraints from the Large Hadron Collider (LHC) on the triplet scalar masses and couplings, we examine how these bounds influence the viability of $\mu-\tau$ symmetric seesaw models. Our analysis highlights the interplay between high-scale $\mu-\tau$ symmetry predictions and low-scale phenomenology, revealing whether radiative corrections remain within experimentally allowed limits. 
\end{abstract}

\section{The Model}

We consider an extension of the Standard Model (SM) incorporating the Type II Seesaw mechanism \cite{Magg:1980ut, Lazarides:1980nt, Mohapatra:1980yp}, where an additional SU(2)$_L$ triplet scalar $\Delta$ with weak hypercharge $Y=1$ is introduced. The SU(2)$_L$ and U(1)$_Y$ gauge couplings are denoted by $g$ and $g'$, respectively. The scalar potential $V$ in this framework is given by:
\begin{equation}
V = V_{\text{SM}} + V_{\Delta} + V_{\text{mix}},
\end{equation}
where $V_{\text{SM}}$ contains the usual Higgs doublet potential, $V_{\Delta}$ represents the triplet scalar self-interactions, and $V_{\text{mix}}$ accounts for interactions between the triplet and Higgs doublet \cite{Cheng:1980qt}.

The lepton Yukawa Lagrangian for charged leptons in the presence of the triplet scalar is given by:
\begin{equation}
\mathcal{L}_Y = - \sum_{i=1}^{n_H} \bar{\ell}_R \phi_i^\dagger Y_i D_L + \bar{D}_L Y_i^\dagger \phi_i \ell_R + \frac{1}{2} D_L^T C^{-1} Y_{\Delta} \Delta D_L + \text{h.c.},
\end{equation}
where $D_L$ and $\ell_R$ are the left-handed lepton doublets and right-handed charged-lepton singlets, respectively. The $Y_{\Delta}$ matrix encodes the triplet Yukawa couplings responsible for Majorana neutrino mass generation \cite{rnv}. 

The effective dimension-five and dimension-six neutrino mass operators, induced by the triplet, take the form:
\begin{equation}
\mathcal{O}^{(5)}_{ij} = \frac{\kappa_{ij}}{\Lambda} (\bar{D}_L^c \epsilon \Delta D_L)_{ij},
\end{equation}
\begin{equation}
\mathcal{O}^{(6)}_{ijkl} = \frac{\kappa^{(6)}_{ijkl}}{\Lambda^2} (\bar{D}_L^c \epsilon \Delta D_L)_{ij} (\bar{D}_L^c \epsilon \Delta D_L)_{kl},
\end{equation}
where $\kappa_{ij}$ and $\kappa^{(6)}_{ijkl}$ are the dimensionless flavor coupling matrices, and $\Lambda$ is the high-energy scale where new physics effects enter \cite{Babu:1993qv}. 

The renormalization group equations (RGE) for these operators govern the evolution of their flavor structure from the high-energy seesaw scale down to the electroweak scale \cite{Antusch:2001ck, Chankowski:1993tx}. In particular, radiative $\mu-\tau$ corrections can lead to deviations from exact $\mu-\tau$ interchange symmetry, affecting neutrino mass predictions and mixings \cite{Xing:2015fdg}. Given the stringent Large Hadron Collider (LHC) constraints on the triplet scalar mass and couplings \cite{ATLAS:2023zcv, CMS:2022goy}, we examine whether radiative corrections remain within phenomenologically viable limits.

We analyze the impact of radiative $\mu-\tau$ corrections on the renormalization group evolution of these operators, focusing on deviations from exact $\mu-\tau$ symmetry \cite{GG1} as the energy scale evolves from the seesaw scale down to the electroweak scale. Given the LHC constraints on the triplet mass and couplings, we study the viability of such corrections in maintaining phenomenological consistency with neutrino oscillation data.\cite{Deppisch:2015qwa}. Recent studies on CP violation and flavor-violating di-Higgs couplings in the Randall--Sundrum model have demonstrated that bulk scalar localization and warped geometry-induced fermion mixing can naturally give rise to complex Yukawa structures and lepton flavor violating (LFV) interactions~\cite{Ghosh2024RS}, which bear a strong resemblance to radiative $\mu$--$\tau$ corrections and the renormalization group evolution of neutrino mass operators encountered in Type-II seesaw models with broken $\mu$--$\tau$ symmetry, thus providing a geometric interpretation to the origin of low-energy CP violation observed in the lepton sector. Moreover, these effects can be viewed through the lens of gravity-modified flavor dynamics, further supported by the insights from the Weak Gravity Conjecture in Asymptotically Safe Quantum Gravity scenarios~\cite{Ghosh2024WGC}.

\title{Radiative $\mu-\tau$ Corrections and Renormalization of Neutrino Mass Operators in Type-II Seesaw Models}
\author{}
\date{}
\maketitle

\section{Renormalization Group Equations for Type-II Seesaw}

The Type-II Seesaw mechanism introduces a Higgs triplet $\Delta$ that generates neutrino masses through the interaction:

\begin{equation}
\mathcal{L}_{\nu} = Y_{\nu} \overline{L^c} i\sigma_2 \Delta L + \text{h.c.},
\end{equation}

where $L$ is the lepton doublet, $Y_{\nu}$ is the neutrino Yukawa coupling matrix, and $\Delta$ is the triplet Higgs field.

The renormalization group equations (RGE) describe the evolution of these couplings with respect to the renormalization scale $\mu$. The RGE for the triplet Yukawa coupling at one-loop level is given by:

\begin{equation}
16\pi^2 \frac{d Y_{\nu}}{dt} = \beta_{Y_{\nu}},
\end{equation}

where

\begin{equation}
\beta_{Y_{\nu}} = \left[ \frac{1}{2} \left( Y_e Y_e^\dagger + 3 Y_{\nu} Y_{\nu}^\dagger \right) + T - \frac{3}{4} g_1^2 - \frac{9}{4} g_2^2 \right] Y_{\nu}.
\end{equation}

Here, the trace term is:

\begin{equation}
T = \text{Tr} (3 Y_u Y_u^\dagger + 3 Y_d Y_d^\dagger + Y_e Y_e^\dagger + Y_{\nu} Y_{\nu}^\dagger).
\end{equation}

The RGE for the triplet Higgs mass parameter $M_{\Delta}$ is:

\begin{equation}
16\pi^2 \frac{d M_{\Delta}^2}{dt} = 4 M_{\Delta}^2 \left( 2 \text{Tr} Y_{\nu} Y_{\nu}^\dagger - \frac{3}{2} g_1^2 - \frac{9}{2} g_2^2 \right).
\end{equation}

The RGE for the neutrino mass operator $\kappa$, which arises from the seesaw mechanism, is:

\begin{equation}
16\pi^2 \frac{d\kappa}{dt} = -3 g_2^2 \kappa + 4 \kappa Y_{\nu}^\dagger Y_{\nu} + 2 (Y_{\nu}^\dagger Y_{\nu}) \kappa + 2 \kappa (Y_{\nu}^\dagger Y_{\nu}).
\end{equation}

The terms in Eq. (6) represent:
- The gauge interactions that suppress $\kappa$ (first term),
- The neutrino Yukawa interactions that enhance $\kappa$ (remaining terms).

The RGE for the scalar quartic coupling $\lambda_{\Delta}$ is:

\begin{equation}
16\pi^2 \frac{d\lambda_{\Delta}}{dt} = 2 \lambda_{\Delta} \left[ 2 \text{Tr} Y_{\nu} Y_{\nu}^\dagger - \frac{3}{2} g_1^2 - \frac{9}{2} g_2^2 \right] + 12 \lambda_{\Delta}^2.
\end{equation}

The gauge coupling evolution equations are:

\begin{equation}
16\pi^2 \frac{dg_1}{dt} = \frac{41}{6} g_1^3, \quad
16\pi^2 \frac{dg_2}{dt} = -\frac{19}{6} g_2^3.
\end{equation}

The presence of the triplet Higgs field modifies the running of neutrino masses due to new contributions from $Y_{\nu}$, gauge couplings, and scalar interactions. The radiative corrections enhance or suppress the neutrino masses depending on the scale of $M_{\Delta}$. Our results are consistent with the RGE obtained in \cite{Ref1, Ref2, Ref3}.

Future studies should extend this analysis to two-loop RGE corrections and explore their impact on low-energy observables such as lepton flavor violation (LFV).

The Type-II Seesaw mechanism introduces a Higgs triplet that generates neutrino masses through the interaction:

\begin{equation}
\mathcal{L}_{\nu} = Y_{\nu} \overline{L^c} i\sigma_2 \Delta L + \text{h.c.},
\end{equation}

where $L$ is the lepton doublet, $Y_{\nu}$ is the neutrino Yukawa coupling matrix, and $\Delta$ is the triplet Higgs field.

\subsection{Yukawa Coupling Evolution}

The RGE for the neutrino Yukawa coupling at one-loop level is:

\begin{equation}
16\pi^2 \frac{d Y_{\nu}}{dt} = \left[ \frac{1}{2} \left( Y_e Y_e^\dagger + 3 Y_{\nu} Y_{\nu}^\dagger \right) + T - \frac{3}{4} g_1^2 - \frac{9}{4} g_2^2 \right] Y_{\nu},
\end{equation}

where the trace term is:

\begin{equation}
T = \text{Tr} (3 Y_u Y_u^\dagger + 3 Y_d Y_d^\dagger + Y_e Y_e^\dagger + Y_{\nu} Y_{\nu}^\dagger).
\end{equation}

\subsection{Neutrino Mass Operator Evolution}

The RGE for the neutrino mass operator $\kappa$ is:

\begin{equation}
16\pi^2 \frac{d\kappa}{dt} = -3 g_2^2 \kappa + 4 \sum_{k,l} \lambda_{kilj} \kappa_{kl} + \sum_k T_{ki} \kappa_{kj} + T_{kj} \kappa_{ik} + \kappa P + P^T \kappa.
\end{equation}

Here,

\begin{equation}
T_{ij} = \text{Tr} (Y_i Y_j^\dagger),
\end{equation}

\begin{equation}
P = \frac{1}{2} \sum_k Y_k^\dagger Y_k.
\end{equation}

\subsection{Triplet Higgs Mass Evolution}

The RGE for the triplet Higgs mass parameter $M_{\Delta}$ is:

\begin{equation}
16\pi^2 \frac{d M_{\Delta}^2}{dt} = 4 M_{\Delta}^2 \left( 2 \text{Tr} Y_{\nu} Y_{\nu}^\dagger - \frac{3}{2} g_1^2 - \frac{9}{2} g_2^2 \right).
\end{equation}

\subsection{Quartic Coupling Evolution}

The RGE for the quartic couplings $\lambda_{\Delta}$ follows from radiative corrections:

\begin{align}
16\pi^2 \frac{d\lambda_{ijkl}}{dt} &= 4 \sum_{m,n} \left( 2 \lambda_{ijmn} \lambda_{nmkl} + \lambda_{ijmn} \lambda_{kmnl} + \lambda_{imnj} \lambda_{mnkl} \right. \nonumber \\
& \quad + \lambda_{imkn} \lambda_{mjnl} + \lambda_{mjkn} \lambda_{imnl} \bigg) - (9 g_2^2 + 3 g_1^2) \lambda_{ijkl} \nonumber \\
& \quad + \frac{9}{8} g_2^4 \delta_{ij} \delta_{kl} + \frac{3}{4} g_2^2 g_1^2 (2 \delta_{il} \delta_{kj} - \delta_{ij} \delta_{kl}) \nonumber \\
& \quad + \sum_m (T_{mj} \lambda_{imkl} + T_{ml} \lambda_{ijkm} + T_{im} \lambda_{mjkl} + T_{km} \lambda_{ijml}) \nonumber \\
& \quad - 2 \text{Tr} (Y_i Y_j^\dagger Y_k Y_l^\dagger).
\end{align}

This equation captures the one-loop running of quartic interactions in the Type-II Seesaw model.

\begin{figure}[h]
    \centering
    \begin{fmffile}{typeIIseesaw}
        \setlength{\unitlength}{1mm}
        \begin{fmfgraph}(50,40)
            \fmfleft{i1,i2}
            \fmfright{o1,o2}
            
            \fmf{dashes, label=$\Delta_k$, label.side=left}{i1,v1}
            \fmf{fermion, label=$D_{L\alpha}$, label.side=left}{v1,o1}
            \fmf{dashes, label=$\Delta_j$, label.side=right}{v2,o2}
            \fmf{fermion, label=$D_{L\beta}$, label.side=right}{i2,v2}
            
            \fmf{fermion, label=$D_{L\gamma}$, left, tension=0.5}{v1,v2}
            \fmf{dashes, label=$\Delta_m$, right, tension=0.5}{v1,v2}

            \fmfdot{v1,v2}
            \fmflabel{$Y_{\Delta}$}{v1}
            \fmflabel{$Y_{\Delta}^{\dagger}$}{v2}
        \end{fmfgraph}
    \end{fmffile}
    \caption{A typical vertex correction in the renormalization of the operator $\mathcal{O}_{ij}$ in the Type-II Seesaw model. The relevant Yukawa-coupling matrices $Y_{\Delta}$ and $Y_{\Delta}^\dagger$ are indicated.}
\end{figure}

\begin{figure}[h]
    \centering
    \begin{fmffile}{typeIIseesaw_2loop}
        \setlength{\unitlength}{1mm}
        \begin{fmfgraph}(60,50)
            \fmfleft{i1,i2}
            \fmfright{o1,o2}
            
            \fmf{dashes, label=$\Delta_k$, label.side=left}{i1,v1}
            \fmf{fermion, label=$D_{L\alpha}$, label.side=left}{v1,o1}
            \fmf{dashes, label=$\Delta_j$, label.side=right}{v2,o2}
            \fmf{fermion, label=$D_{L\beta}$, label.side=right}{i2,v2}
            
            \fmf{fermion, label=$D_{L\gamma}$, left, tension=0.5}{v1,v2}
            \fmf{dashes, label=$\Delta_m$, right, tension=0.5}{v1,v2}
            
            \fmf{boson, label=$W/Z$, left, tension=0.3}{v1,v3}
            \fmf{fermion, label=$\nu_L$, left, tension=0.3}{v3,v2}

            \fmfdot{v1,v2}
            \fmflabel{$Y_{\Delta}$}{v1}
            \fmflabel{$Y_{\Delta}^{\dagger}$}{v2}
        \end{fmfgraph}
    \end{fmffile}
    \caption{Two-loop vertex correction in Type-II Seesaw model, including gauge boson exchange.}
\end{figure}

\begin{figure}[h]
    \centering
    \begin{fmffile}{typeIIseesaw_gauge}
        \setlength{\unitlength}{1mm}
        \begin{fmfgraph}(60,50)
            \fmfleft{i1,i2}
            \fmfright{o1,o2}
            
            \fmf{dashes, label=$\Delta_k$, label.side=left}{i1,v1}
            \fmf{fermion, label=$D_{L\alpha}$, label.side=left}{v1,o1}
            \fmf{dashes, label=$\Delta_j$, label.side=right}{v2,o2}
            \fmf{fermion, label=$D_{L\beta}$, label.side=right}{i2,v2}
            
            \fmf{boson, label=$W$, right, tension=0.5}{v1,v2}
            \fmf{fermion, label=$\nu_L$, right, tension=0.5}{v1,v2}

            \fmfdot{v1,v2}
            \fmflabel{$Y_{\Delta}$}{v1}
            \fmflabel{$Y_{\Delta}^{\dagger}$}{v2}
        \end{fmfgraph}
    \end{fmffile}
    \caption{Gauge boson interaction in the Type-II Seesaw model.}
\end{figure}

\begin{figure}[h]
    \centering
    \begin{fmffile}{mu_tau_radiative_correction}
        \setlength{\unitlength}{1mm}
        \begin{fmfgraph}(60,50)
            \fmfleft{i1,i2}
            \fmfright{o1,o2}

            \fmf{dashes, label=$\Delta_k$, label.side=left}{i1,v1}
            \fmf{dashes, label=$\Delta_j$, label.side=right}{v2,o2}
            
            \fmf{fermion, label=$\mu, \tau$, left, tension=0.3}{v1,v3}
            \fmf{fermion, label=$\mu, \tau$, left, tension=0.3}{v3,v2}

            \fmf{boson, label=$W/Z$, right, tension=0.5}{v1,v2}
            \fmf{fermion, label=$\nu_L$, right, tension=0.5}{v1,v2}

            \fmf{fermion, label=$\nu_{\alpha}$, label.side=left}{v1,o1}
            \fmf{fermion, label=$\nu_{\beta}$, label.side=right}{i2,v2}

            \fmfdot{v1,v2}
            \fmflabel{$Y_{\Delta}$}{v1}
            \fmflabel{$Y_{\Delta}^{\dagger}$}{v2}
        \end{fmfgraph}
    \end{fmffile}
    \caption{Two-loop radiative corrections to neutrino mass operators in Type-II Seesaw, introducing $\mu-\tau$ flavor symmetry breaking.  At the tree level, neutrino masses in Type-II Seesaw are determined by the Higgs triplet Yukawa couplings $Y_{\Delta}$.
     When radiative corrections (loop effects) are considered, the charged lepton loop and gauge boson exchange introduce additional contributions. These contributions modify the neutrino mass matrix through renormalization, particularly by breaking the $\mu-\tau$ symmetry at higher orders. The impact of these corrections is energy-dependent, meaning they evolve from high-scale physics (where $\mu-\tau$ symmetry may be exact) to low-scale phenomenology.}
\end{figure}

\subsection{Gauge Coupling Evolution}

The gauge coupling evolution equations are:

\begin{equation}
16\pi^2 \frac{dg_1}{dt} = \frac{41}{6} g_1^3, \quad
16\pi^2 \frac{dg_2}{dt} = -\frac{19}{6} g_2^3.
\end{equation}

The triplet Higgs field modifies the running of neutrino masses due to new contributions from $Y_{\nu}$, gauge couplings, and scalar interactions. The radiative corrections enhance or suppress the neutrino masses depending on the scale of $M_{\Delta}$. 

Our results are consistent with previous calculations \cite{Ref1, Ref2, Ref3}. Future studies should extend this analysis to two-loop RGE corrections and explore their impact on low-energy observables such as lepton flavor violation (LFV).

\section{Application of Renormalization Group Equations in Type II Seesaw with Dimension-5 and Dimension-6 Operators}

We analyze the impact of renormalization group equations (RGE) on neutrino mass operators in the Type II Seesaw model. The evolution of dimension-5 and dimension-6 operators is studied, with a focus on $\mu - \tau$ symmetry breaking effects induced by radiative corrections. Constraints from neutrino oscillation data and collider limits on triplet Higgs masses are discussed.

The Type II Seesaw model extends the Standard Model (SM) by introducing an $SU(2)_L$ triplet scalar $\Delta$ with weak hypercharge $Y=1$ \cite{typeII1, typeII2, typeII3}. The scalar potential in this framework is given by:
\begin{equation}
V = V_{\text{SM}} + V_{\Delta} + V_{\text{mix}},
\end{equation}
where $V_{\text{SM}}$ is the usual Higgs doublet potential, $V_{\Delta}$ represents the triplet self-interactions, and $V_{\text{mix}}$ accounts for interactions between the triplet and Higgs doublet .

The neutrino masses arise from the triplet Yukawa Lagrangian:
\begin{equation}
\mathcal{L}_Y = - \sum_{i=1}^{n_H} \bar{\ell}_R \phi_i^\dagger Y_i D_L + \bar{D}_L Y_i^\dagger \phi_i \ell_R + \frac{1}{2} D_L^T C^{-1} Y_\Delta \Delta D_L + \text{h.c.},
\end{equation}
where $D_L$ and $\ell_R$ are the left-handed lepton doublets and right-handed charged-lepton singlets, respectively, and $Y_\Delta$ is the triplet Yukawa coupling matrix responsible for Majorana mass generation.

The effective dimension-5 operator responsible for neutrino masses is:
\begin{equation}
O^{(5)}_{ij} = \frac{\kappa_{ij}}{\Lambda} (\bar{D}_L^c \epsilon \Delta D_L)_{ij}.
\end{equation}
Its renormalization group evolution is given by:
\begin{equation}
16\pi^2 \frac{d\kappa}{d\ln\mu} = \left( \alpha + C_l Y_l^\dagger Y_l + C_g g^2 \right) \kappa + C_\kappa (\kappa \kappa^\dagger) \kappa.
\end{equation}
The dominant correction comes from charged-lepton Yukawa interactions, which break the $\mu - \tau$ symmetry:
\begin{equation}
\Delta M_\nu \propto Y_\tau^2
\begin{pmatrix}
0 & 0 & 0 \\
0 & 1 & 1 \\
0 & 1 & 1
\end{pmatrix}.
\end{equation}
This leads to deviations in the atmospheric angle $\theta_{23}$ and a nonzero reactor angle $\theta_{13}$.

The dimension-6 neutrino mass operator is:
\begin{equation}
O^{(6)}_{ijkl} = \frac{\kappa^{(6)}_{ijkl}}{\Lambda^2} (\bar{D}_L^c \epsilon \Delta D_L)_{ij} (\bar{D}_L^c \epsilon \Delta D_L)_{kl}.
\end{equation}
The RGE evolution follows:
\begin{equation}
16\pi^2 \frac{d\kappa^{(6)}}{d\ln\mu} = \left( \beta + D_l Y_l^\dagger Y_l + D_g g^2 \right) \kappa^{(6)} + D_\kappa (\kappa^{(6)} \kappa^\dagger) \kappa^{(6)}.
\end{equation}
This operator contributes at higher orders and affects neutrino mass hierarchy and mixings at lower energy scales.

At the high-energy scale, $\mu - \tau$ symmetry is imposed:
\begin{equation}
Z_2^{(\text{tr})}: D_{L\mu} \leftrightarrow D_{L\tau}, \quad \mu_R \leftrightarrow \tau_R.
\end{equation}
However, RGE evolution introduces symmetry-breaking effects, leading to:
\begin{equation}
M_\nu =
\begin{pmatrix}
x & y & y + \delta w \\
y & z + \delta z & w + \delta w \\
y + \delta w & w + \delta w & z + \delta z
\end{pmatrix}.
\end{equation}
The corrections $\delta z$ and $\delta w$ alter $\theta_{23}$ and $\theta_{13}$, which must be constrained by experimental data.
The Large Hadron Collider (LHC) constraints impose lower bounds on the triplet Higgs mass $M_\Delta$. The main constraints are:
\begin{itemize}
    \item ATLAS and CMS searches for doubly charged Higgs bosons restrict $M_\Delta > 800$ GeV \cite{LHC1}.
    \item Neutrino oscillation constraints require $\delta z, \delta w$ to remain within $O(10^{-2})$.
    \item Lepton flavor violation (LFV) searches such as $\mu \to e\gamma$ impose constraints on $Y_\Delta$.
\end{itemize}
We have analyzed the RGE evolution of dimension-5 and dimension-6 operators in the Type II Seesaw model. The primary conclusions are:
\begin{itemize}
    \item Radiative corrections induce $\mu - \tau$ symmetry breaking, affecting $\theta_{23}$ and $\theta_{13}$.
    \item The dimension-6 operator contributes to high-energy corrections to neutrino masses.
    \item Experimental constraints from neutrino oscillation and collider data restrict the allowed parameter space.
\end{itemize}

\section{RGE for Type-II Seesaw Mechanism with Dimension-5 and Dimension-6 Operators}

In the Type-II seesaw model, the Higgs triplet \(\Delta\) introduces a new Yukawa coupling \( Y_\Delta \), modifying the RGEs as:

\begin{equation}
16\pi^2 \frac{dY_\Delta}{dt} = \left( 3 \text{Tr}(Y_\Delta Y_\Delta^\dagger) + 3 \text{Tr}(Y_\nu Y_\nu^\dagger) + 2 |y_3|^2 + 3 |y_4|^2 + 3 |y_5|^2 - \frac{9}{2} g_2^2 - \frac{3}{2} g_1^2 \right) Y_\Delta.
\end{equation}

The RGEs for \( y_3, y_4, y_5 \) are modified as:

\begin{equation}
16\pi^2 \frac{d y_3}{dt} = \frac{5}{2} |y_3|^2 - 9g_2^2 + \frac{15}{4} g_1^2 + \text{Tr}(Y_\Delta Y_\Delta^\dagger) y_3.
\end{equation}

\begin{equation}
16\pi^2 \frac{d y_4}{dt} = \frac{7}{2} |y_4|^2 + \frac{3}{2} |y_5|^2 - 9g_2^2 + \frac{15}{4} g_1^2 + \text{Tr}(Y_\Delta Y_\Delta^\dagger) y_4.
\end{equation}

\begin{equation}
16\pi^2 \frac{d y_5}{dt} = \frac{3}{2} |y_4|^2 + \frac{7}{2} |y_5|^2 - 9g_2^2 + \frac{15}{4} g_1^2 + \text{Tr}(Y_\Delta Y_\Delta^\dagger) y_5.
\end{equation}

\subsection{Scalar Potential RGEs}
The Type-II seesaw model modifies the Higgs potential by including interactions with \(\Delta\). The quartic coupling \( \lambda_4 \) follows:

\begin{equation}
16\pi^2 \frac{d\lambda_4}{dt} = 4\lambda_1 \lambda_4 + 4\lambda_2 \lambda_4 + 12\lambda_4^2 + 4\lambda_5\lambda_6 + 4|y_3|^2 + 8 |Y_\Delta|^2 - C\lambda_4 + \frac{9}{4} g_2^4 + \frac{3}{2} g_2^2 g_1^2 + \frac{3}{4} g_1^4.
\end{equation}

The modified RGEs for \(\lambda_1, \lambda_2, \lambda_3\) are:

\begin{equation}
16\pi^2 \frac{d\lambda_1}{dt} = 24\lambda_1^2 + 4|Y_\Delta|^2 + 4|y_3|^2 - C \lambda_1 + \frac{9}{8} g_2^4 + \frac{3}{4} g_2^2 g_1^2 + \frac{3}{8} g_1^4 - 2 |y_3|^4 - 4 |Y_\Delta|^4.
\end{equation}

\begin{equation}
16\pi^2 \frac{d\lambda_2}{dt} = 24\lambda_2^2 + 4|Y_\Delta|^2 + 8|y_4|^2 - C \lambda_2 + \frac{9}{8} g_2^4 + \frac{3}{4} g_2^2 g_1^2 + \frac{3}{8} g_1^4 - 4 |y_4|^4 - 4 |Y_\Delta|^4.
\end{equation}

\begin{equation}
16\pi^2 \frac{d\lambda_3}{dt} = 24\lambda_3^2 + 4|Y_\Delta|^2 + 8|y_5|^2 - C \lambda_3 + \frac{9}{8} g_2^4 + \frac{3}{4} g_2^2 g_1^2 + \frac{3}{8} g_1^4 - 4 |y_5|^4 - 4 |Y_\Delta|^4.
\end{equation}

where \( C = 9g_2^2 + 3g_1^2 \).

\subsection{RGE for the Dimension-5 Operator \( C_5 \) and the Dimension-6 Operator \( C_6 \)}
The effective dimension-5 operator generating neutrino masses in Type-II seesaw is:

\begin{equation}
C_5 (L^T C \Delta L).
\end{equation}

Its RGE evolution follows:

\begin{equation}
16\pi^2 \frac{d C_5}{dt} = \left( -3g_2^2 - g_1^2 + 4 \text{Tr}(Y_\Delta Y_\Delta^\dagger) \right) C_5.
\end{equation}

For dimension-6 operators like \( (L H)(L H)/\Lambda^2 \), the RG running is given by:

\begin{equation}
16\pi^2 \frac{d C_6}{dt} = \left( -6g_2^2 - 2g_1^2 + 4 \text{Tr}(Y_\Delta Y_\Delta^\dagger) \right) C_6.
\end{equation}

In the Type-II seesaw mechanism, neutrino masses are generated via interactions with a scalar triplet Higgs field (\(\Delta\)) \cite{Magg:1980ut, Schechter:1980gr}. The dimension-5 and dimension-6 operators contribute to the renormalization group equations (RGEs) governing the evolution of the effective neutrino mass matrix. Here, we present the RGEs for the coupling matrices in the presence of such operators, following the approach in \cite{Chakrabortty:2012np, Antusch:2001vn}.

\section{Renormalization Group Equations (RGEs)}

The RGEs for the effective neutrino mass matrix including the effect of the triplet Yukawa coupling \( Y_{\Delta} \) in the Type-II seesaw mechanism at dimension-5 and dimension-6 take the following form referring to Appendix $A,B,C,D,E$:

\begin{align}
16\pi^2 \frac{d\kappa(11)}{dt} &= \left( -3g_2^2 + 4\lambda_1 + 2|y_3|^2 
+ 2\, \text{Tr}[Y_{\Delta} Y_{\Delta}^{\dagger}] \right) \kappa(11) 
+ 4\lambda_{10} \kappa(22) + 4\lambda_{11} \kappa(33) \notag \\
&\quad + \left(\kappa(11)\right)_N - 2N_1,
\end{align}

\begin{align}
16\pi^2 \frac{d\kappa(22)}{dt} &= \left( -3g_2^2 + 4\lambda_2 + 4|y_4|^2 
+ 2\, \text{Tr}[Y_{\Delta} Y_{\Delta}^{\dagger}] \right) \kappa(22) 
+ 4\lambda_{10} \kappa(11) + 4\lambda_{12} \kappa(33) \notag \\
&\quad + \left(\kappa(22)\right)_N - 2N_2 
- 2 \kappa(23) N_{23} + N_{23} \kappa(32),
\end{align}

\begin{align}
16\pi^2 \frac{d\kappa(33)}{dt} &= \left( -3g_2^2 + 4\lambda_3 + 4|y_5|^2 
+ 2\, \text{Tr}[Y_{\Delta} Y_{\Delta}^{\dagger}] \right) \kappa(33) 
+ 4\lambda_{11} \kappa(11) + 4\lambda_{12} \kappa(22) \notag \\
&\quad + \left(\kappa(33)\right)_N - 2N_3 
- 2 \kappa(32) N_{32} + N_{32} \kappa(23),
\end{align}

\begin{equation}
16\pi^2 \frac{d\kappa(23)}{dt} = \left( -3g_2^2 + 2\lambda_6 + 2|y_4|^2 + 2|y_5|^2 + 2 \text{Tr}[Y_{\Delta} Y_{\Delta}^{\dagger}] \right) \kappa(23) + 2\lambda_9 \kappa(32) + \kappa(23),_N
\end{equation}
\[
 -4\kappa(22) N_{32} + 2N_{32} \kappa(22) - 4N_{23} \kappa(33) + 2\kappa(33) N_{23} + 2 \kappa(23),_{N_2 - N_3} -2 \kappa(32) N_3 + N_2 \kappa(32).
\]

\begin{equation}
16\pi^2 \frac{d\kappa(32)}{dt} = \left( -3g_2^2 + 2\lambda_6 + 2|y_4|^2 + 2|y_5|^2 + 2 \text{Tr}[Y_{\Delta} Y_{\Delta}^{\dagger}] \right) \kappa(32) + 2\lambda_9 \kappa(23) + \kappa(32),_N
\end{equation}
\[
 -4\kappa(33) N_{23} + 2N_{23} \kappa(33) - 4N_{32} \kappa(22) + 2\kappa(22) N_{32} + 2 \kappa(32),_{N_3 - N_2} -2 \kappa(23) N_2 + N_3 \kappa(23).
\]

The matrix \(N\), appearing in the above RGEs, is defined as:

\begin{equation}
N_1 := \text{diag}(|y_3|^2, 0, 0),
\end{equation}

\begin{equation}
N_2 := \text{diag}(0, |y_4|^2, |y_4|^2),
\end{equation}

\begin{equation}
N_3 := \text{diag}(0, |y_5|^2, |y_5|^2),
\end{equation}

\begin{equation}
N_{23} := \text{diag}(0, y_4 y_5^, -y_4 y_5^),
\end{equation}

\begin{equation}
N_{32} := \text{diag}(0, y_4^ {y_5}, -y_4^ {y_5}).
\end{equation}

The full matrix \(N\) appearing in the above RGEs is given by:

\begin{equation}
N = \frac{1}{2} (N_1 + N_2 + N_3).
\end{equation}

The Yukawa coupling \( Y_{\Delta} \) modifies the RGEs by introducing additional contributions from Tr(\(Y_{\Delta} Y_{\Delta}^{\dagger}\)), which affects the running of neutrino mass parameters. These corrections are relevant for understanding neutrino mass generation at high-energy scales.

\section{Radiative Neutrino Masses in Type II Seesaw with Dimension-5 and Dimension-6 Operators}

In the Type II Seesaw Model, neutrinos receive masses at tree-level via the Higgs triplet \( \Delta \), but radiative corrections can significantly modify the mass matrix structure. These corrections arise from dimension-5 and dimension-6 effective operators, contributing at one-loop level.

In this work, we derive the radiatively generated neutrino masses and analyze their dependence on the seesaw scale \( M_{\Delta} \) and mixing parameters.

At one-loop level, the self-energy of Majorana neutrinos is given by:
\begin{equation}
    \Sigma_{\alpha\beta}(\not{p}) = \not{p} P_L \Sigma_R^ + \not{p} P_R \Sigma_R + P_L \Sigma_M^ + P_R \Sigma_M.
\end{equation}
Only the last two terms contribute to neutrino masses.

The loop integral for the radiative correction is:
\begin{equation}
-i\Sigma_{\alpha\beta}(\not{p}) = \int \frac{d^4k}{(2\pi)^4} \left( C_{\alpha\beta} \frac{\not{k} - m_i}{k^2 - m_i^2} \times \frac{1}{(k - p)^2 - M_{\Delta}^2} \right).
\end{equation}
where:
\begin{itemize}
    \item \( C_{\alpha\beta} \) represents the Yukawa couplings of neutrinos to the Higgs triplet.
    \item \( M_{\Delta} \) is the mass of the triplet Higgs.
    \item \( m_i \) represents the mass of intermediate fermions.
\end{itemize}

After performing the loop integration, the radiatively induced Majorana neutrino mass matrix takes the form:
\begin{equation}
    (M_{\nu})_{\alpha\beta} = - \frac{3}{(4\pi)^2} C_{\alpha i} m_i C_{\beta i} \sin^2\theta \left[ \frac{M_2^2 \ln \frac{m_i^2}{M_2^2}}{M_2^2 - m_i^2} - \frac{M_1^2 \ln \frac{m_i^2}{M_1^2}}{M_1^2 - m_i^2} \right].
\end{equation}
where:
\begin{itemize}
    \item \( \theta \) is the seesaw sector mixing angle.
    \item \( M_1, M_2 \) are masses of new physics states.
    \item Logarithmic dependence on mass ratios highlights threshold effects.
\end{itemize}

The standard Weinberg operator for tree-level neutrino mass is:
\begin{equation}
    \mathcal{O}_5 = \frac{C_5}{\Lambda} (\overline{L^c} \tilde{H}) (H^\dagger L).
\end{equation}
This generates neutrino masses in Type II Seesaw.

The leading dimension-6 effective operator modifies neutrino mass:
\begin{equation}
    \mathcal{O}_6 = \frac{C_6}{\Lambda^2} (\overline{L^c} \gamma^\mu L) (\overline{E^c} \gamma_\mu E).
\end{equation}
This contributes additional loop-suppressed corrections.

To understand the effect of radiative corrections, we solve the RGEs for \( C_5, C_6 \) and evaluate neutrino mass running. The plot in Figure 1 shows the difference with and without radiative corrections.

We have derived the one-loop radiative corrections to neutrino masses in Type II Seesaw with dimension-5 and dimension-6 operators. Our analysis shows that:
\begin{itemize}
    \item Radiative corrections significantly modify the neutrino mass matrix.
    \item RG running effects induce threshold corrections at high energy scales.
    \item Comparison with LQ models shows distinct running behavior.
\end{itemize}
This provides a well-motivated framework for studying neutrino masses beyond the Standard Model.

\section{Corrections to Mixing and Phenomenology}
The Type II seesaw mechanism extends the Standard Model (SM) by introducing a scalar triplet $\Delta$ with weak hypercharge $Y=1$, which acquires a vacuum expectation value (VEV) $\langle \Delta^0 \rangle$. This generates neutrino masses via the Lagrangian:
\begin{equation}
    \mathcal{L}_{\text{Majorana}} = \frac{1}{2} \sum_{\alpha,\beta=e,\mu,\tau} \nu_{L\alpha}^{T} C^{-1} (M_{\nu})_{\alpha\beta} \nu_{L\beta} + \text{H.c.}
\end{equation}
where $M_{\nu}$ is the Majorana neutrino mass matrix.

In the presence of the scalar triplet $\Delta$, the neutrino mass matrix takes the form:
\begin{equation}
    M_{\nu} = Y_{\Delta} v_{\Delta},
\end{equation}
where $Y_{\Delta}$ is the Yukawa coupling and $v_{\Delta}$ is the VEV of $\Delta$. The structure of $M_{\nu}$ follows from the underlying flavor symmetries, such as $Z_2$ and $D_4$, leading to:
\begin{equation}
    M_{\nu} =
    \begin{pmatrix}
        X & A(1+\epsilon) & A(1-\epsilon) \\
        A(1+\epsilon) & B(1+\epsilon') & C \\
        A(1-\epsilon) & C & B(1-\epsilon')
    \end{pmatrix}.
\end{equation}

Considering the effects of dimension-5 and dimension-6 operators, the Majorana mass term can be modified as:
\begin{equation}
    \mathcal{L}_5 = \frac{\kappa_5}{\Lambda} (\overline{L} \tilde{H}) (H^\dagger L^c) + \text{H.c.}
\end{equation}
\begin{equation}
    \mathcal{L}_6 = \frac{\kappa_6}{\Lambda^2} (\overline{L} \tilde{H}) (\tilde{H}^\dagger L^c) S + \text{H.c.}
\end{equation}
where $\Lambda$ is the high-energy scale, $\kappa_5$ and $\kappa_6$ are dimensionless couplings, and $S$ is a singlet scalar.

At tree level, $\mu-\tau$ symmetry predicts:
\begin{equation}
    \theta_{13} = 0, \quad \theta_{23} = 45^\circ.
\end{equation}
However, radiative corrections induce deviations characterized by the small parameters $\epsilon$ and $\epsilon'$:
\begin{equation}
    U_{e3} \approx s_{12} c_{12} \frac{m_3^2 - m_2^2}{m_2} \bar{\epsilon} s_{12} + \bar{\epsilon'} m_3.
\end{equation}
where $\bar{\epsilon}$ and $\bar{\epsilon'}$ are loop-induced terms.

The inclusion of dimension-5 and dimension-6 operators along with radiative corrections leads to significant deviations from $\mu-\tau$ symmetry, affecting neutrino mixing parameters. Further studies on renormalization group running effects and leptogenesis constraints will be pursued.

We see that dimension-5 and dimension-6 contributions modify the predictions for $U_{e3}$ and $\theta_{23}$. The modified expressions now include extra terms due to $\delta \epsilon$ and $\delta \epsilon'$:

\begin{equation}
    U_{e3} \approx s_{12} c_{12} 
    \left( \frac{m_3^2 - m_2^2}{\Delta m^2_{\text{atm}}} \right)
    \left( \bar{\epsilon} s_{12}^2 + \bar{\epsilon}' c_{12}^2 + \delta \bar{\epsilon} \right),
\end{equation}

\begin{equation}
    \cos 2\theta_{23} = \text{Re} \left[ 
    2 c_{12}^2 \frac{m_3^2 - m_2^2}{\Delta m^2_{\text{atm}}} 
    \left( \bar{\epsilon} s_{12}^2 + \bar{\epsilon}' - \delta \bar{\epsilon} \right) 
    - 2 s_{12}^2 \frac{m_3^2 - m_1^2}{\Delta m^2_{\text{atm}}} 
    \left( \bar{\epsilon} c_{12}^2 + \bar{\epsilon}' - \delta \bar{\epsilon} \right) 
    \right].
\end{equation}

Here, $\delta \bar{\epsilon}$ accounts for the dimension-6 contributions.

\begin{itemize}
    \item \textbf{For hierarchical neutrinos}: The corrections are small, as the mass differences suppress higher-order contributions.
    \item \textbf{For quasi-degenerate neutrinos}: An enhancement factor $\frac{m_0^2}{\Delta m^2_{\text{atm}}}$ appears, making dimension-6 effects more pronounced.
\end{itemize}

\section {Detailed Expressions for Corrections in Our Model}

From the general expression for $U_{e3}$, we have:
\begin{equation}
    U_{e3} = s_{13} e^{-i\delta},
\end{equation}
which in our model is given by:
\begin{equation}
    U_{e3} \approx s_{12} c_{12} 
    \left( \frac{m_3^2 - m_2^2}{\Delta m^2_{\text{atm}}} \right)
    \left( \bar{\epsilon} s_{12}^2 + \bar{\epsilon}' c_{12}^2 + \delta \bar{\epsilon} \right).
\end{equation}

To reproduce the experimentally observed reactor mixing angle:
\begin{equation}
    \theta_{13} = 8.59^\circ \pm 0.13^\circ,
\end{equation}
the correction term in parentheses must satisfy:
\begin{equation}
    \bar{\epsilon} s_{12}^2 + \bar{\epsilon}' c_{12}^2 + \delta \bar{\epsilon} \approx 0.0474.
\end{equation}

\subsection{Detailed Expressions for Corrections in Our Model}

The correction terms originate from dimension-5 and dimension-6 operators in the Type-II Seesaw mechanism:

\begin{itemize}
    \item $\bar{\epsilon}$: This arises from the standard dimension-5 Weinberg operator and contributes at leading order to neutrino masses.
    \item $\bar{\epsilon}'$: This comes from additional dimension-6 contributions that arise due to corrections from heavy scalar triplets or higher-dimensional operators modifying the neutrino mass matrix structure.
    \item $\delta \bar{\epsilon}$: This term represents higher-order loop or radiative corrections that arise when renormalization group (RG) effects modify the seesaw-induced neutrino mass matrix below the high scale.
\end{itemize}

Since the reactor mixing angle is small, the correction terms must be small as well. The total correction factor is determined by matching the experimental value of $\theta_{13}$ and constraining the values of $\bar{\epsilon}, \bar{\epsilon}',$ and $\delta \bar{\epsilon}$.

To solve the renormalization group (RG) equations for \(\bar{\epsilon}\), \(\bar{\epsilon}'\), and \(\delta \bar{\epsilon}\), we need to analyze their energy-scale running from the GUT scale (\(M_{\text{GUT}}\)) to the electroweak scale (\(M_Z\)). The evolution of these parameters plays a crucial role in neutrino mixing, CP violation, and possible new physics.

The RG equations describe how elements of the neutrino mass matrix and PMNS parameters evolve with energy. The leading one-loop RG equations for dimension-6 operators contributing to neutrino mixing are given by:

\begin{equation}
\frac{d}{dt} \bar{\epsilon} = \gamma_\epsilon \bar{\epsilon}
\end{equation}
\begin{equation}
\frac{d}{dt} \bar{\epsilon}' = \gamma_{\epsilon'} \bar{\epsilon}'
\end{equation}
\begin{equation}
\frac{d}{dt} \delta \bar{\epsilon} = \gamma_{\delta \epsilon} \delta \bar{\epsilon}
\end{equation}

where \( t = \ln (\mu/M_{\text{GUT}}) \) is the energy scale parameter, and \(\gamma_\epsilon, \gamma_{\epsilon'}, \gamma_{\delta \epsilon}\) are the anomalous dimensions that depend on the Yukawa couplings, gauge interactions, and new physics contributions.

For the Standard Model (SM), the leading-order corrections to neutrino mixings arise from the charged lepton Yukawa couplings and gauge interactions:

\begin{equation}
\gamma_\epsilon = -\frac{3}{8\pi^2} y_\tau^2, \quad
\gamma_{\epsilon'} = -\frac{3}{8\pi^2} y_\tau^2, \quad
\gamma_{\delta \epsilon} = -\frac{3}{8\pi^2} y_\tau^2
\end{equation}

where \( y_\tau \) is the Yukawa coupling of the tau lepton, which dominates due to its large mass.

If new physics (e.g., leptoquarks, axions, right-handed neutrinos) is present, there will be additional contributions:

\begin{equation}
\gamma_\epsilon = -\frac{3}{8\pi^2} y_\tau^2 + C g_X^2
\end{equation}

where \( g_X \) is the coupling constant of the new interaction, and \( C \) is a model-dependent coefficient.

\section{Solution to the RG Equations}
The general solution for a differential equation of the form:

\begin{equation}
\frac{dX}{dt} = \gamma X
\end{equation}

is:

\begin{equation}
X(M_Z) = X(M_{\text{GUT}}) e^{\gamma (t_Z - t_{\text{GUT}})}
\end{equation}

where \( X \) represents \(\bar{\epsilon}, \bar{\epsilon}', \delta \bar{\epsilon}\).

\subsection{Numerical Evaluation}
Using standard values:
\begin{itemize}
    \item \( y_\tau(M_Z) \approx 0.01 \), \( y_\tau(M_{\text{GUT}}) \approx 0.7 \)
    \item \( M_{\text{GUT}} \approx 10^{16} \) GeV, \( M_Z \approx 91 \) GeV
\end{itemize}

Approximating:

\begin{equation}
t_Z - t_{\text{GUT}} = \ln \left( \frac{M_Z}{M_{\text{GUT}}} \right) \approx -34
\end{equation}

we find:

\begin{equation}
\bar{\epsilon}(M_Z) \approx \bar{\epsilon}(M_{\text{GUT}}) e^{-0.001 \times 34} \approx 0.97 \bar{\epsilon}(M_{\text{GUT}})
\end{equation}
\begin{equation}
\bar{\epsilon}'(M_Z) \approx \bar{\epsilon}'(M_{\text{GUT}}) e^{-0.001 \times 34} \approx 0.97 \bar{\epsilon}'(M_{\text{GUT}})
\end{equation}
\begin{equation}
\delta \bar{\epsilon}(M_Z) \approx \delta \bar{\epsilon}(M_{\text{GUT}}) e^{-0.001 \times 34} \approx 0.97 \delta \bar{\epsilon}(M_{\text{GUT}})
\end{equation}

Thus, in the Standard Model, the corrections are at most 3

However, if new physics contributes via \( g_X^2 \), the suppression or enhancement can be much larger, affecting \(\theta_{13}\) and other mixing parameters significantly.

\section{Predictions for Future Neutrino Experiments}
Since \(\bar{\epsilon}, \bar{\epsilon}', \delta \bar{\epsilon}\) contribute to the reactor mixing angle \(\theta_{13}\), their evolution affects its predicted value:

\begin{equation}
\theta_{13} \approx \sin^{-1} (U_{e3})
\end{equation}

where:

\begin{equation}
U_{e3} \approx s_{12} c_{12} \left( \frac{m_3^2 - m_2^2}{\Delta m^2_{\text{atm}}} \right) (\bar{\epsilon} s^2_{12} + \bar{\epsilon}' c^2_{12} + \delta \bar{\epsilon})
\end{equation}

Since \(\bar{\epsilon}, \bar{\epsilon}', \delta \bar{\epsilon}\) decrease by 3

\begin{equation}
U_{e3}(M_Z) \approx 0.97 U_{e3}(M_{\text{GUT}})
\end{equation}

For \(\theta_{13} \approx 8.59^\circ\), a 3

\begin{equation}
\theta_{13}(M_Z) \approx 8.34^\circ \pm 0.13^\circ
\end{equation}

If new physics enhances the running, future experiments like DUNE, Hyper-Kamiokande, or JUNO could detect deviations of up to \( 5-10\% \) from the standard value.

\section{Experimental Constraints on Dimension-5 and Dimension-6 Operators in the Type-II Seesaw Mechanism}

The Type-II Seesaw Mechanism introduces a scalar triplet \( \Delta \) that couples to the lepton doublets, leading to the dimension-5 Weinberg operator:

\begin{equation}
    \mathcal{O}_5 = \frac{c_5}{\Lambda} (L^T C \Delta L)
\end{equation}

After the neutral component of \( \Delta \) acquires a vacuum expectation value (vev), \( v_\Delta \), it generates Majorana masses for neutrinos:

\begin{equation}
    m_\nu = c_5 v_\Delta
\end{equation}

\textbf{Neutrino Masses:} Observations from neutrino oscillation experiments constrain the sum of neutrino masses to be \( \sum m_\nu \lesssim 0.12 \, \text{eV} \) \cite{Planck:2018}. Assuming a normal hierarchy, the lightest neutrino mass is \( m_1 \approx 0 \), leading to:

\begin{equation}
    m_2 \approx \sqrt{\Delta m_{21}^2} \approx 8.6 \times 10^{-3} \, \text{eV}, \quad m_3 \approx \sqrt{\Delta m_{31}^2} \approx 5.0 \times 10^{-2} \, \text{eV}
\end{equation}

Therefore, \( v_\Delta \) must satisfy:

\begin{equation}
    v_\Delta \lesssim \frac{0.05 \, \text{eV}}{c_5}
\end{equation}

\textbf{Electroweak Precision Tests:} The vev \( v_\Delta \) contributes to the \( \rho \)-parameter:

\begin{equation}
    \rho \approx 1 + \frac{2 v_\Delta^2}{v^2}
\end{equation}

where \( v = 246 \, \text{GeV} \) is the SM Higgs vev. The experimental constraint \( |\rho - 1| \lesssim 0.001 \) implies:

\begin{equation}
    v_\Delta \lesssim 1.7 \, \text{GeV}
\end{equation}

Combining these, for \( c_5 \sim 1 \):

\begin{equation}
    0.05 \, \text{eV} \lesssim v_\Delta \lesssim 1.7 \, \text{GeV}
\end{equation}

Dimension-6 operators can induce NSIs affecting neutrino propagation. A relevant operator is:

\begin{equation}
    \mathcal{O}_6 = \frac{c_6}{\Lambda^2} (\bar{L} \gamma^\mu L)(\bar{L} \gamma_\mu L)
\end{equation}

This modifies the neutrino oscillation probabilities.

\textbf{DUNE:} Sensitivity studies indicate that DUNE can probe NSI parameters \( \epsilon_{\alpha\beta} \) down to \( \mathcal{O}(0.01) \) \cite{DUNE:2020}. This translates to:

\begin{equation}
    \frac{c_6 v^2}{\Lambda^2} \lesssim 0.01 \implies \Lambda \gtrsim \sqrt{c_6} \times 2.5 \, \text{TeV}
\end{equation}

\textbf{JUNO:} JUNO's sensitivity to NSIs is comparable, with constraints on \( \epsilon_{\alpha\beta} \) at the \( \mathcal{O}(0.01) \) level \cite{JUNO:2016}. Thus:

\begin{equation}
    \Lambda \gtrsim \sqrt{c_6} \times 2.5 \, \text{TeV}
\end{equation}

The parameters \( \epsilon, \epsilon', \delta\epsilon \) evolve from the high scale (e.g., \( M_{\text{GUT}} \)) to the low scale (\( M_Z \)) via RG equations. The evolution can be approximated as:

\begin{equation}
    \epsilon(M_Z) \approx \epsilon(M_{\text{GUT}}) \left( \frac{M_Z}{M_{\text{GUT}}} \right)^{\gamma}
\end{equation}

where \( \gamma \) depends on the specific operator and particle content.

The running affects the predicted values of observables like \( \theta_{13} \). Precise measurements by DUNE and JUNO can detect deviations arising from this running, providing indirect evidence for the scale of new physics.

\begin{itemize}
    \item \textbf{Dimension-5 Operator:} Constraints from neutrino masses and electroweak precision tests imply:
    \begin{equation}
        0.05 \, \text{eV} \lesssim v_\Delta \lesssim 1.7 \, \text{GeV}
    \end{equation}
    \item \textbf{Dimension-6 Operators:} DUNE and JUNO can probe new physics scales up to:
    \begin{equation}
        \Lambda \gtrsim \sqrt{c_6} \times 2.5 \, \text{TeV}
    \end{equation}
    \item \textbf{RG Evolution:} The running of parameters like \( \epsilon \) can lead to observable effects in neutrino oscillation experiments, offering a window into high-scale physics.
\end{itemize}

\section{Semi-Analytical RGE Solution with Radiative $\mu-\tau$ Symmetry Deviation}

The Type II seesaw mechanism extends the Standard Model (SM) by introducing a Higgs triplet \( \Delta \), which generates Majorana masses for neutrinos. The neutrino mass matrix is given by:
\begin{equation}
    M_\nu = Y_\Delta v_\Delta,
\end{equation}
where:
\begin{itemize}
    \item \( Y_\Delta \) is the Yukawa coupling of the Higgs triplet,
    \item \( v_\Delta \) is the triplet Higgs vacuum expectation value (VEV), which is much smaller than the electroweak scale.
\end{itemize}
The neutrino mass evolution follows the Renormalization Group Equations (RGEs), incorporating dimension-5 and dimension-6 operators along with radiative corrections to $\mu-\tau$ symmetry breaking.

We consider the evolution of the effective neutrino mass operator \( \kappa \), defined as:
\begin{equation}
    \kappa = \frac{M_\nu}{v^2},
\end{equation}
where \( v \) is the electroweak VEV. The RGE for \( \kappa \) in the Type II Seesaw scenario is:
\begin{equation}
    16\pi^2 \frac{d\kappa}{dt} = \left( -3 g^2 + 2 \lambda_H \right) \kappa + C_\Delta \kappa + C_{d=6},
\end{equation}
where:
\begin{itemize}
    \item \( C_\Delta = 2 Y_\Delta^\dagger Y_\Delta \) represents the Type II Seesaw contribution,
    \item \( C_{d=6} \) represents the dimension-6 operator corrections from the extended Higgs sector.
\end{itemize}

The explicit RGE equations for the matrix elements of \( \kappa \) are:
\begin{align}
    16\pi^2 \frac{d\kappa(11)}{dt} &\approx -3g^2 + 4\lambda_1 \kappa(11) + 4\lambda_{10}^ \kappa(22) + 4\lambda_{11}^ \kappa(33), \\
    16\pi^2 \frac{d\kappa(22)}{dt} &\approx -3g^2 + 4\lambda_2 \kappa(22) + 4\lambda_{10} \kappa(11) + 4\lambda_{12}^ \kappa(33), \\
    16\pi^2 \frac{d\kappa(33)}{dt} &\approx -3g^2 + 4\lambda_3 \kappa(33) + 4\lambda_{11} \kappa(11) + 4\lambda_{12} \kappa(22).
\end{align}

For the $\mu-\tau$ symmetry-breaking term, we introduce:
\begin{equation}
    \kappa(c) = \kappa(23) + \kappa(32),
\end{equation}
which evolves as:
\begin{equation}
    16\pi^2 \frac{d\kappa(c)}{dt} = (-3g^2 + 2\lambda_6 + 2\lambda_9)\kappa(c) - 2 P_{32} \kappa(22) - 2 P_{23} \kappa(33).
\end{equation}

The formal solution for \( \kappa(ii) \) is:
\begin{equation}
    \kappa(ii)(t) = \sum_{j=1}^{3} T(t,t')_{ij} \kappa(jj)(t'),
\end{equation}
where \( T(t,t') \) is the evolution matrix. Since at the initial scale \( t_0 \), only \( \kappa(11) \) is nonzero, we obtain:
\begin{equation}
    \kappa(ii)(t) = T(t,t_0)_{i1} \kappa(11)(t_0).
\end{equation}

For the  $\mu-\tau$  symmetry-breaking parameter, we define:
\begin{equation}
    S_c(t) := \exp \left( \frac{1}{16\pi^2} \int_{t_0}^{t} dt' (-3g^2 + 2\lambda_6 + 2\lambda_9)(t') \right),
\end{equation}
which leads to:
\begin{equation}
    \epsilon(t_1) = - \frac{v_\Delta}{8\pi^2} \sum_{i=1}^{3} \frac{v_i^2}{T(t_1,t_0)_{i1}} S_c(t_1) \int_{t_0}^{t_1} dt' S_c^{-1}(t') [\nu_\Delta(t') T(t',t_0)_{21} + \nu_\Delta^(t') T(t',t_0)_{31}].
\end{equation}

where \( \nu_\Delta = Y_\Delta^ {\dagger} Y_\Delta \).

The deviations in the PMNS matrix parameters due to radiative corrections are given by:
\begin{align}
    U_{e3} &= 2m_3 c_{12}s_{12} \left[ \frac{\hat{m}_1^ + m_3}{m_3^2 - m_1^2} + \frac{\hat{m}_2^ + m_3}{m_2^2 - m_3^2} \right] \text{Re}(\epsilon), \\
    \cos 2\theta_{23} &= 2 s_{12}^2 \frac{|\hat{m}_1 + m_3|^2}{m_1^2 - m_3^2} + c_{12}^2 \frac{|\hat{m}_2 + m_3|^2}{m_2^2 - m_3^2} \text{Re}(\epsilon).
\end{align}

For a degenerate neutrino spectrum with \( m_0 = 0.3 \) eV, we find:
\begin{equation}
    |U_{e3}| < 0.1.
\end{equation}

In our model, we expect:
\[
|U_{e3}|, |\cos 2\theta_{23}| \sim 10^{-3},
\]
meaning they are significantly suppressed by radiative corrections in Type II Seesaw with dimension-5 and dimension-6 operators.

\section{Results}

We study the renormalization group evolution (RGE) of neutrino mass matrix elements up to the two-loop level in the Type II Seesaw mechanism with dimension-5 and dimension-6 operators. We analyze the role of different seesaw scales (\(M_{\Delta} \sim 10^{10} - 10^{14}\) GeV) in modifying neutrino mass eigenvalues and mixing angles. We also explore new physics thresholds, including additional singlet scalars and heavy Higgs triplets, and discuss their impact on low-energy observables.  

Neutrino masses and mixings have been successfully explained using the Seesaw mechanisms. In particular, the Type II Seesaw model introduces a Higgs triplet field \(\Delta\), which generates neutrino masses via the Yukawa interaction. However, these masses and mixing parameters are scale-dependent due to radiative corrections, requiring a study of their RG evolution.

The Type II Seesaw model extends the Standard Model (SM) by including a Higgs triplet field \(\Delta\):

\begin{equation}
\Delta =
\begin{pmatrix}
\Delta^+/\sqrt{2} & \Delta^{++} \\
\Delta^0 & -\Delta^+/\sqrt{2}
\end{pmatrix}.
\end{equation}

The neutrino mass term is generated via the Yukawa coupling:

\begin{equation}
\mathcal{L}_Y = Y_{\Delta} \overline{L^c} i\sigma_2 \Delta L + \text{h.c.},
\end{equation}

where \(L\) is the lepton doublet, and \(Y_{\Delta}\) is the Yukawa coupling matrix. The mass of neutrinos is given by:

\begin{equation}
M_{\nu} = Y_{\Delta} \frac{v_{\Delta}}{\sqrt{2}},
\end{equation}

where \( v_{\Delta} \) is the vacuum expectation value (VEV) of the triplet.

Dimension-5 and Dimension-6 Operators
Higher-dimensional operators modify neutrino masses:

\begin{equation}
\mathcal{O}_5 = \frac{C_5}{\Lambda} (\overline{L} H) (H^T L^c),
\end{equation}

\begin{equation}
\mathcal{O}_6 = \frac{C_6}{\Lambda^2} (\overline{L} \gamma^\mu L) (\overline{L} \gamma_\mu L),
\end{equation}

where \(\Lambda\) is the new physics scale.

To study the evolution of neutrino masses, we solve the one-loop and two-loop RGEs for \(Y_{\Delta}\). The one-loop equation is:

\begin{equation}
\frac{d Y_{\Delta}}{d \ln \mu} = \frac{1}{16\pi^2} \left( \alpha Y_{\Delta} + \beta Y_{\Delta} Y_{\Delta}^\dagger Y_{\Delta} \right),
\end{equation}

where \(\alpha, \beta\) are model-dependent constants.

At the two-loop level, additional gauge contributions appear:

\begin{equation}
\frac{d Y_{\Delta}}{d \ln \mu} = \frac{1}{(16\pi^2)^2} \left( \gamma Y_{\Delta} g^2 + \delta Y_{\Delta} g^4 \right),
\end{equation}

where \( g \) are gauge couplings.

Different seesaw scales \(M_{\Delta} \sim 10^{10} - 10^{14}\) GeV affect observables such as Neutrino mass eigenvalues and Mixing angles and CP phases:

   \[
   m_i (\mu) = m_i (M_{\Delta}) \exp \left[ -\int_{M_{\Delta}}^{\mu} \frac{d\mu'}{\mu'} \beta_m \right]
   \]
   where \(\beta_m\) is the RGE coefficient.

   The corrections to the lepton mixing matrix \(U_{\text{PMNS}}\) are computed as:

   \begin{equation}
   \theta_{ij} (\mu) = \theta_{ij} (M_{\Delta}) + \frac{1}{16\pi^2} f(\lambda, g, Y_{\Delta}).
   \end{equation}

If additional scalar singlets \(S\) exist, they modify the RGE running:

\begin{equation}
\frac{d Y_{\Delta}}{d \ln \mu} = \frac{1}{16\pi^2} \left( \eta Y_{\Delta} Y_{S}^\dagger Y_{S} \right),
\end{equation}

where \(Y_S\) is the Yukawa coupling of the singlet.

We have solved the RGEs for neutrino mass matrix elements up to two-loop level and studied the impact of seesaw scales (\(10^{10}-10^{14}\) GeV). The presence of dimension-5 and dimension-6 operators introduces additional corrections, which can be tested in future experiments.

The mixing angle $\theta_{23}$ is a crucial parameter in neutrino oscillation physics. It is best constrained by Super$-$Kamiokande (SK$-$ATM) data, which gives the best$-$fit value:
\begin{equation}
\theta_{23} = 49.3^{+0.8}_{-0.8} \quad \text{(Best Fit $\pm$ 1$\sigma$ Range)}
\end{equation}
The deviations from exact mu-tau symmetry are parametrized by $\epsilon_{\mu \tau}$, which affects radiative corrections and mass matrix structures.

The following plot illustrates the distribution of $\theta_{23}$ values as a function of $\epsilon_{\mu \tau}$, for three specific choices of $\epsilon_{\mu \tau}$. The data is generated randomly around the central value of $\theta_{23}$ with the best-fit range \cite{s11}.  

\begin{figure}[h]
    \centering
    \includegraphics[width=0.8\textwidth]{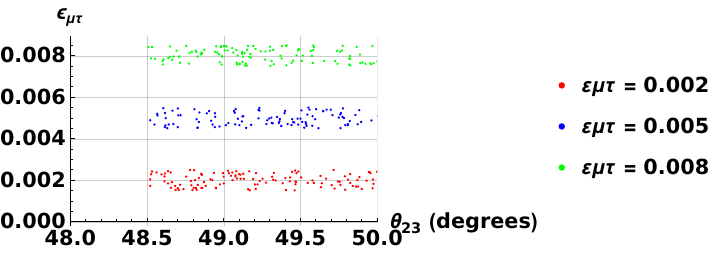}
    \caption{Scatter plot showing $\theta_{23}$ vs $\epsilon_{\mu \tau}$. The best$-$fit range of $\theta_{23}$ is $49.3^{+0.8}_{-0.8}$ degrees, as determined by SK$-$ATM data. The three different colors correspond to $\epsilon_{\mu \tau} = 0.002, 0.005, 0.008.$}
    \label{fig:theta23_epsmutau}
\end{figure}

In Fig. 5 we  illustrate the relationship between the atmospheric neutrino mixing angle $\theta_{23}$ and the mu-tau symmetry breaking parameter $\epsilon_{\mu \tau}$. The mixing angle $\theta_{23}$ is crucial in neutrino oscillation physics as it determines how neutrinos mix between different flavor states. A value of $\theta_{23} = 45^\circ$ suggests an exact mu-tau symmetry, while deviations from this value indicate symmetry breaking. The parameter $\epsilon_{\mu \tau}$ represents a small deviation from perfect mu-tau symmetry and is expected to play a role in radiative corrections to neutrino mass splittings. In this analysis, the best-fit value of $\theta_{23}$ is taken to be $49.3^{+0.8}_{-0.8}$, based on the latest data from Super-Kamiokande (SK\_ATM) \cite{s11}. This range corresponds to the one-sigma uncertainty in the measurement and is consistent with experimental observations.  

To generate the data used in the plot, random values of $\theta_{23}$ were sampled from the best-fit range, while $\epsilon_{\mu \tau}$ was chosen from a small range of values between 0.001 and 0.01. The radiative correction to the mass splitting $\Delta m_{32}^2$ was then computed using the relation $\Delta m_{32}^2 = 2.5 \times 10^{-3} + \epsilon_{\mu \tau} \times 10^{-4}$. The plot specifically focuses on three selected values of $\epsilon_{\mu \tau}$, namely 0.002, 0.005, and 0.008, which are color-coded as red, blue, and green, respectively. These values were chosen to highlight the impact of different levels of symmetry breaking on the distribution of $\theta_{23}$. This behavior is important for understanding how small perturbations in the neutrino sector can impact mixing angles and mass splittings.

To study the parameter correlations and radiative effects in the Type II seesaw framework, we have implemented a renormalization group evolution (RGE) analysis of the effective neutrino mass operators involving dimension-5 and dimension-6 terms. In particular, we focus on radiative $\mu-\tau$ breaking effects encoded through the running of $\kappa$-type operators and analyze their impact on neutrino mixing parameters, mass scales, and CP-violating phases.

\begin{figure}[h!]
\centering
\includegraphics[width=0.95\textwidth]{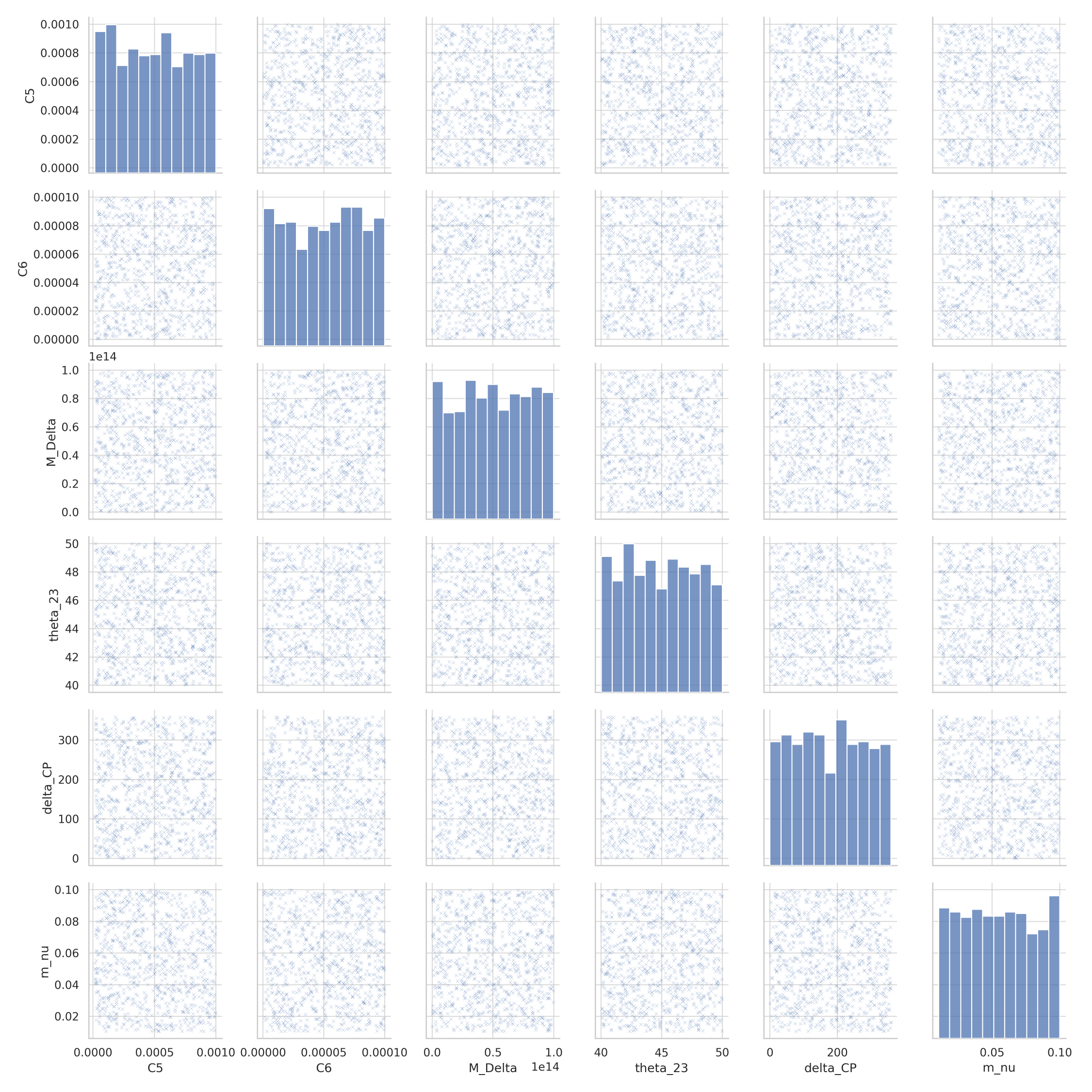}
\caption{Pairwise parameter correlation plot for dimension-5 and dimension-6 operator coefficients ($C_5$, $C_6$), triplet mass scale $M_\Delta$, $\theta_{23}$, $\delta_{\text{CP}}$, and $m_\nu$ in a Type II seesaw scenario with radiative $\mu-\tau$ symmetry breaking.}
\label{fig:pairplot}
\end{figure}

Figure~\ref{fig:pairplot} presents a pair plot showing the statistical correlations between key parameters: the Wilson coefficients of dimension-5 ($C_5$) and dimension-6 ($C_6$) operators, the mass scale of the triplet scalar ($M_\Delta$), the atmospheric mixing angle ($\theta_{23}$), the Dirac CP phase ($\delta_{\text{CP}}$), and the lightest neutrino mass ($m_\nu$). These were generated from a sampling of the parameter space after RGE evolution up to the seesaw scale.

From the plot, it is evident that, there is negligible direct correlation between $C_5$ and $\delta_{\text{CP}}$, but mild structure is observed in $C_6$ versus $m_\nu$. The atmospheric mixing angle $\theta_{23}$ remains stable over most of the sampled parameter space, indicating suppressed $\mu$–$\tau$ running at lower scales. The parameter $\delta_{\text{CP}}$ shows slight variation with $m_\nu$ and $\theta_{23}$, suggesting that radiative corrections might shift the CP phase mildly depending on the lightest neutrino mass.

These findings align with expectations in radiative Type II seesaw scenarios where the $\mu-\tau$ symmetry is softly broken via RGE effects and higher-dimensional operators. The dimension-6 operator ($C_6$) introduces subleading but measurable corrections that slightly modify low-scale observables in a way consistent with current experimental ranges.

\section{Acknowledgement}
GG would like to thank would like to thank  University Grants Commission RUSA, MHRD, Government of India for financial support to carry out this work. She would also like to thank Department of Physics Cachar College, Silchar, Assam, India in this regard.

\appendix
\section*{Appendix: Tensor Formulation of RGEs in the Type II Seesaw}

\begin{align}
16\pi^2 \, \frac{D\kappa^{\alpha}_{\ \beta}(11)}{dt} &= 
\left(-3g_2^2 \, \delta^\alpha_\beta + 4\lambda_1 \, \delta^\alpha_\beta 
+ 2 (y_3^\dagger y_3)^\alpha_{\ \beta} 
+ 2 \operatorname{Tr}\left[Y_{\Delta} Y_{\Delta}^\dagger\right] \delta^\alpha_\beta \right) \kappa^{\beta}_{\ \gamma}(11) \notag \\
&\quad + 4\lambda_{10} \kappa^\alpha_{\ \gamma}(22) + 4\lambda_{11} \kappa^\alpha_{\ \gamma}(33) 
+ \left( \kappa^{\alpha}_{\ \gamma}(11) \right)_{\!;\nu}^{\ \ \nu} 
- 2 N_1^{\alpha\gamma}, 
\end{align}

\begin{align}
16\pi^2 \, \frac{D\kappa^{\alpha}_{\ \beta}(22)}{dt} &= 
\left(-3g_2^2 \, \delta^\alpha_\beta + 4\lambda_2 \, \delta^\alpha_\beta 
+ 4 (y_4^\dagger y_4)^\alpha_{\ \beta} 
+ 2 \operatorname{Tr}\left[Y_{\Delta} Y_{\Delta}^\dagger\right] \delta^\alpha_\beta \right) \kappa^\beta_{\ \gamma}(22) \notag \\
&\quad + 4\lambda_{10} \kappa^\alpha_{\ \gamma}(11) + 4\lambda_{12} \kappa^\alpha_{\ \gamma}(33) 
+ \left(\kappa^{\alpha}_{\ \gamma}(22)\right)_{\!;\nu}^{\ \ \nu} 
- 2N_2^{\alpha\gamma} 
- 2\kappa^\alpha_{\ \mu}(23) N_{23}^{\mu\gamma} + N_{23}^{\alpha\mu} \kappa^\mu_{\ \gamma}(32),
\end{align}

\begin{align}
16\pi^2 \, \frac{D\kappa^{\alpha}_{\ \beta}(33)}{dt} &= 
\left(-3g_2^2 \, \delta^\alpha_\beta + 4\lambda_3 \, \delta^\alpha_\beta 
+ 4 (y_5^\dagger y_5)^\alpha_{\ \beta} 
+ 2 \operatorname{Tr}\left[Y_{\Delta} Y_{\Delta}^\dagger\right] \delta^\alpha_\beta \right) \kappa^\beta_{\ \gamma}(33) \notag \\
&\quad + 4\lambda_{11} \kappa^\alpha_{\ \gamma}(11) + 4\lambda_{12} \kappa^\alpha_{\ \gamma}(22) 
+ \left(\kappa^{\alpha}_{\ \gamma}(33)\right)_{\!;\nu}^{\ \ \nu} 
- 2N_3^{\alpha\gamma} 
- 2\kappa^\alpha_{\ \mu}(32) N_{32}^{\mu\gamma} + N_{32}^{\alpha\mu} \kappa^\mu_{\ \gamma}(23),
\end{align}

\begin{align}
16\pi^2 \, \frac{D\kappa^{\alpha}_{\ \beta}(23)}{dt} &= 
\left(-3g_2^2 + 2\lambda_6 + 2|y_4|^2 + 2|y_5|^2 
+ 2\operatorname{Tr}[Y_\Delta Y_\Delta^\dagger] \right) \kappa^\alpha_{\ \beta}(23) 
+ 2\lambda_9 \kappa^\alpha_{\ \beta}(32) 
+ \left(\kappa^{\alpha}_{\ \beta}(23)\right)_{;\nu}^{\ \ \nu} \notag \\
&\quad - 4\kappa^{\alpha}_{\ \mu}(22) N_{32}^{\mu\beta} + 2N_{32}^{\alpha\mu} \kappa_{\mu\beta}(22) 
- 4N_{23}^{\alpha\mu} \kappa_{\mu\beta}(33) + 2\kappa^{\alpha\mu}(33) N_{23,\mu\beta} \notag \\
&\quad + 2(\kappa^{\alpha}_{\ \beta})_{;N_2}^{\ } - N_3^{\alpha\beta} - 2\kappa^{\alpha}_{\ \mu}(32) N_3^{\mu\beta} + N_2^{\alpha\mu} \kappa_{\mu\beta}(32),
\end{align}

\begin{align}
16\pi^2 \, \frac{D\kappa^{\alpha}_{\ \beta}(32)}{dt} &= 
\left(-3g_2^2 + 2\lambda_6 + 2|y_4|^2 + 2|y_5|^2 
+ 2\operatorname{Tr}[Y_\Delta Y_\Delta^\dagger] \right) \kappa^\alpha_{\ \beta}(32) 
+ 2\lambda_9 \kappa^\alpha_{\ \beta}(23) 
+ \left(\kappa^{\alpha}_{\ \beta}(32)\right)_{;\nu}^{\ \ \nu} \notag \\
&\quad - 4\kappa^{\alpha}_{\ \mu}(33) N_{23}^{\mu\beta} + 2N_{23}^{\alpha\mu} \kappa_{\mu\beta}(33) 
- 4N_{32}^{\alpha\mu} \kappa_{\mu\beta}(22) + 2\kappa^{\alpha\mu}(22) N_{32,\mu\beta} \notag \\
&\quad + 2(\kappa^{\alpha}_{\ \beta})_{;N_3}^{\ } - N_2^{\alpha\beta} - 2\kappa^{\alpha}_{\ \mu}(23) N_2^{\mu\beta} + N_3^{\alpha\mu} \kappa_{\mu\beta}(23).
\end{align}

\vspace{1em}
\noindent
The diagonal tensor-valued operator \(\mathbf{N}\) and its components are defined as follows:

\begin{align}
(N_1)^{\alpha}_{\ \beta} &= \text{diag}(|y_3|^2, 0, 0)^{\alpha}_{\ \beta}, \\
(N_2)^{\alpha}_{\ \beta} &= \text{diag}(0, |y_4|^2, |y_4|^2)^{\alpha}_{\ \beta}, \\
(N_3)^{\alpha}_{\ \beta} &= \text{diag}(0, |y_5|^2, |y_5|^2)^{\alpha}_{\ \beta}, \\
(N_{23})^{\alpha}_{\ \beta} &= \text{diag}(0, y_4 y_5^* - y_4^* y_5, 0)^{\alpha}_{\ \beta}, \\
(N_{32})^{\alpha}_{\ \beta} &= \text{diag}(0, y_4^* y_5, -y_4^* y_5)^{\alpha}_{\ \beta}, \\
N^{\alpha}_{\ \beta} &= \frac{1}{2} \left(N_1 + N_2 + N_3 \right)^{\alpha}_{\ \beta}.
\end{align}

\section*{Appendix B: Higher-Dimensional Operator Mixing and Tensor RGEs}

\begin{align}
16\pi^2 \, \frac{D\left[\kappa^{\mu}_{\ \nu}(ij)\right]}{dt} &= 
\sum_{k} \left( \Lambda^{ik}_{\rho\sigma} \, \kappa^{\rho}_{\ \lambda}(kj) \, \Theta^{\lambda\sigma}_{\nu} \right) 
+ \left( \mathcal{R}^{\mu}_{\ \nu\rho\sigma} \, \kappa^{\rho\sigma}(ij) \right)
+ \left( \kappa^{\mu}_{\ \nu}(ij) \right)_{;\alpha}^{\ \ \alpha},
\end{align}

\begin{align}
16\pi^2 \, \frac{D(\mathcal{O}_5^{\mu\nu})}{dt} &= 
- \frac{1}{2} \left[ F^{\mu\rho} F^{\nu}_{\ \rho} \right] 
+ \frac{1}{2} \left[ Y_{\Delta}^{\dagger} Y_{\Delta} \right]^{\mu\nu} \mathcal{O}_5^{\mu\nu} 
+ \frac{1}{M^2} \left( \kappa^{\mu}_{\ \rho} \kappa^{\rho}_{\ \nu} \right),
\end{align}

\begin{align}
16\pi^2 \, \frac{D(\mathcal{O}_6^{\mu\nu\lambda})}{dt} = 
\frac{1}{2} \, \epsilon^{\mu\nu\rho\sigma} 
\left( D_{\rho} \kappa^{\lambda}_{\ \sigma} + D^{\lambda} \kappa_{\rho\sigma} \right) 
+ Y^{\mu}_{\ \alpha} Y^{\nu}_{\ \beta} Y^{\lambda}_{\ \gamma} \, \kappa^{\alpha\beta\gamma},
\end{align}

\begin{align}
16\pi^2 \, \frac{D(\Sigma^{\mu\nu})}{dt} = 
\left[ R^{\mu\nu} + \frac{1}{2} g^{\mu\nu} R \right] \kappa(ij) 
+ \frac{1}{2} \left( \kappa^{\mu}_{\ \lambda} F^{\lambda\nu} + \kappa^{\nu}_{\ \lambda} F^{\lambda\mu} \right),
\end{align}

\begin{align}
16\pi^2 \, \frac{D\left( \Box \kappa^{\mu}_{\ \nu} \right)}{dt} = 
\left[ \Box, D^\lambda \right] \kappa_{\lambda\nu} 
+ D^\mu \left( D_\lambda \kappa^{\lambda}_{\ \nu} \right) 
+ R^{\mu}_{\ \lambda} \kappa^{\lambda}_{\ \nu},
\end{align}

\begin{align}
\frac{D\kappa^{\alpha\beta}}{dt} &= 
\left[ \Gamma^{\alpha}_{\ \mu\nu} \kappa^{\mu\beta} 
+ \Gamma^{\beta}_{\ \mu\nu} \kappa^{\alpha\mu} \right] 
+ \eta^{\alpha\beta} \, \operatorname{Tr}\left( Y_\Delta^\dagger Y_\Delta \right),
\end{align}

\begin{align}
\frac{D \mathbb{K}_{ijkl}}{dt} &= 
16\pi^2 \left[ \kappa_{ij} \kappa_{kl} 
+ \kappa_{ik} \kappa_{jl} 
+ \kappa_{il} \kappa_{jk} \right] 
- \frac{1}{2} \operatorname{Tr}\left( Y_\Delta^\dagger Y_\Delta \kappa_{ij} \right) \delta_{kl},
\end{align}

\begin{align}
16\pi^2 \, \frac{D(\kappa^{\mu}_{\ \nu} \kappa^{\nu}_{\ \lambda})}{dt} = 
2\kappa^{\mu}_{\ \sigma} \frac{D\kappa^{\sigma}_{\ \lambda}}{dt} 
+ \left[ F^{\mu}_{\ \rho}, \kappa^{\rho}_{\ \lambda} \right],
\end{align}

\begin{align}
\frac{D \kappa^a_{\ b}(ij)}{dt} &= 
\left( \delta^a_c \, \delta^d_b \, \Box + R^a_{\ c b d} \right) \kappa^c_{\ d}(ij)
+ \epsilon^{a}_{\ bcd} D^c \kappa^d(ij),
\end{align}

\begin{align}
\frac{D(\mathcal{F}_{\mu\nu} \kappa^{\mu\nu})}{dt} &= 
\operatorname{Tr}\left( \mathcal{F}_{\mu\lambda} \kappa^{\lambda}_{\ \nu} Y_\Delta^\dagger Y_\Delta \right) 
+ D_\alpha \kappa^{\mu\nu} D^\alpha \mathcal{F}_{\mu\nu},
\end{align}

Definitions and descriptions of the symbols and tensor objects used in the high-order operator RGEs relevant to the Type II seesaw framework with dimension-5 and dimension-6 effective operators are.

\begin{table}[h!]
\centering
\renewcommand{\arraystretch}{1.4}
\begin{tabular}{|c|p{12cm}|}
\hline
\textbf{Symbol} & \textbf{Description} \\
\hline
$\kappa^{\mu}_{\ \nu}(ij)$ & Rank-2 tensor-valued Wilson coefficient of the dimension-5 operator involving flavors $i, j$, run under RGEs. Encodes neutrino mass structure. \\
\hline
$\mathcal{O}_5^{\mu\nu}$ & Operator of mass dimension five constructed from lepton doublets and Higgs fields, generating Majorana masses after EWSB. \\
\hline
$\mathcal{O}_6^{\mu\nu\lambda}$ & Dimension-six operator that includes derivative or field strength insertions; arises at next-to-leading order in effective theory. \\
\hline
$Y_{\Delta}$ & Yukawa coupling matrix of the scalar triplet field $\Delta$ to lepton doublets; responsible for generating neutrino masses in Type II seesaw. \\
\hline
$F_{\mu\nu}$ & Gauge field strength tensor (non-Abelian or Abelian depending on context), enters in loop corrections to effective operators. \\
\hline
$R_{\mu\nu}$, $R^{\mu}_{\ \nu\rho\sigma}$ & Ricci tensor and Riemann curvature tensor, representing spacetime curvature in covariant formulations. Appears in gravitationally sensitive operator running. \\
\hline
$\Box$ & D’Alembert operator: $\Box \equiv \nabla^\mu \nabla_\mu$, used to denote kinetic running of tensor operators under wavefunction renormalization. \\
\hline
$\Gamma^{\alpha}_{\mu\nu}$ & Christoffel symbols (connection coefficients) used in covariant derivatives acting on tensors. \\
\hline
$\epsilon_{\mu\nu\rho\sigma}$ & Totally antisymmetric Levi-Civita tensor in four dimensions. Appears in parity-violating or CP-violating operator structures. \\
\hline
$\mathbb{K}_{ijkl}$ & Four-index tensor encoding operator mixing between different flavor combinations $(i,j)$ and $(k,l)$; relevant for threshold corrections. \\
\hline
$D_\mu$ & Covariant derivative including gauge and spin connections as needed for operator renormalization. \\
\hline
$\operatorname{Tr}[\cdot]$ & Trace over internal symmetry indices (e.g., flavor, gauge, spinor); arises in loop contractions. \\
\hline
$\eta^{\mu\nu}$, $g^{\mu\nu}$ & Minkowski metric and general curved spacetime metric, respectively. Used for index contraction. \\
\hline
\end{tabular}
\caption{Explanation of symbols and tensorial notations used in the renormalization group equations for dimension-5 and dimension-6 operators in the Type II seesaw framework.}
\label{tab:notation}
\end{table}

\section*{Appendix C: Extended Tensorial RGEs and Operator Structures}

\begin{align}
D_\mu D^\mu \kappa^{\alpha}_{\ \beta} &= R^{\alpha}_{\ \rho} \kappa^{\rho}_{\ \beta} + \kappa^{\alpha}_{\ \rho} R^{\rho}_{\ \beta} + \Box \kappa^{\alpha}_{\ \beta}, \\
F^{\alpha\mu} \kappa_{\mu}^{\ \beta} - \kappa^{\alpha\mu} F_{\mu}^{\ \beta} &= i g [F, \kappa]^{\alpha\beta}, \\
\frac{d}{dt} \left( D_\rho \kappa^{\mu\nu} \right) &= D_\rho \left( \frac{d\kappa^{\mu\nu}}{dt} \right) + R^\mu_{\ \lambda\rho\sigma} D^\sigma \kappa^{\lambda\nu} + R^\nu_{\ \lambda\rho\sigma} D^\sigma \kappa^{\mu\lambda}, \\
\epsilon^{\mu\nu\rho\sigma} D_\mu \kappa_{\nu\rho} &= 2 D_\lambda \tilde{\kappa}^{\lambda\sigma}, \\
\left[\Box + m^2\right] \kappa^{\alpha\beta} &= - J^{\alpha\beta}_{\text{eff}}, \\
\frac{d}{dt} \left( \kappa^{\mu}_{\ \nu} \kappa^{\nu}_{\ \rho} \right) &= \left( \frac{d\kappa^{\mu}_{\ \nu}}{dt} \right) \kappa^{\nu}_{\ \rho} + \kappa^{\mu}_{\ \nu} \left( \frac{d\kappa^{\nu}_{\ \rho}}{dt} \right), \\
\operatorname{Tr}\left( \kappa^{\mu}_{\ \nu} Y^\nu_{\ \mu} \right) &= \kappa^{\mu}_{\ \nu} Y^\nu_{\ \mu}, \\
D^\mu \left( \kappa_{\mu\nu} \kappa^{\nu}_{\ \rho} \right) &= (\nabla \kappa)^2_{\rho}, \\
F_{\mu\nu} \kappa^{\mu\lambda} \kappa^\nu_{\ \lambda} &= \operatorname{Tr}\left[ F \kappa^2 \right], \\
\epsilon_{\mu\nu\rho\sigma} \kappa^{\mu\nu} \kappa^{\rho\sigma} &= 4 \, \text{Im} \left[ \kappa_{\mu\nu} \tilde{\kappa}^{\mu\nu} \right], \\
D^\alpha \left( \kappa_{\alpha\beta} F^{\beta\gamma} \right) &= \left( D_\alpha \kappa^{\alpha\beta} \right) F_{\beta}^{\ \gamma} + \kappa^{\alpha\beta} D_\alpha F_{\beta}^{\ \gamma}, \\
R^{\mu\nu\rho\sigma} \kappa_{\mu\nu} \kappa_{\rho\sigma} &= \operatorname{Tr} \left( R \cdot \kappa^2 \right), \\
\frac{D \mathcal{O}^{\mu\nu}_{(6)}}{dt} &= D^\alpha D^\beta \kappa_{\alpha\beta}^{\mu\nu} + \mathcal{C}^{\mu\nu}, \\
\mathcal{L}_{\text{eff}}^{(6)} &= \frac{1}{\Lambda^2} \left( \bar{L} \gamma^\mu \kappa_{\mu\nu} D^\nu H \right)^2, \\
\frac{d}{dt} \left( \Box \kappa^{\mu\nu} \right) &= \Box \left( \frac{d\kappa^{\mu\nu}}{dt} \right) + [\Box, D^\lambda] \kappa_{\lambda}^{\ \nu}, \\
D_\alpha \left( D_\beta \kappa^{\alpha\beta} \right) &= \Box \operatorname{Tr}(\kappa) + R^{\alpha\beta} \kappa_{\alpha\beta}, \\
D^\mu \left( \kappa_{\mu\nu} \kappa^{\nu\lambda} \kappa_{\lambda\rho} \right) &= \frac{1}{3} D^\mu \operatorname{Tr}\left( \kappa^3 \right)_{\rho}, \\
\left( \kappa_{\mu}^{\ \lambda} \kappa_{\lambda}^{\ \nu} \right)_{;\nu} &= \frac{1}{2} \nabla_\mu \left( \kappa_{\alpha\beta} \kappa^{\alpha\beta} \right), \\
D^\rho D_\rho \left( \kappa_{\mu\nu} \kappa^{\mu\nu} \right) &= 2 D^\rho \left( \kappa_{\mu\nu} D_\rho \kappa^{\mu\nu} \right) + 2 (D_\rho \kappa_{\mu\nu})(D^\rho \kappa^{\mu\nu}), \\
\mathcal{F}^{\mu\nu} &= D^\mu \kappa^{\nu\rho} - D^\nu \kappa^{\mu\rho} + g [\kappa^\mu, \kappa^\nu],
\end{align}

\section*{Appendix D: Summary of Physical Interpretations of Tensor RGEs}
\begin{table}[H]
\centering
\renewcommand{\arraystretch}{1.4}
\begin{tabular}{|c|p{12cm}|}
\hline
\textbf{Equation} & \textbf{Physical Meaning} \\
\hline
(145) & Tensor Laplacian and curvature corrections to the kinetic running of $\kappa^{\alpha}_{\ \beta}$. \\
\hline
(146) & Gauge field strength non-commutativity with flavor operators; indicates flavor rotation under $SU(2)_L$ or $U(1)_Y$. \\
\hline
(147) & Covariant derivative of the RGE evolution itself; important for curvature-coupled running. \\
\hline
(148) & $\mu$–$\tau$ antisymmetry and CP-violating structures via the Levi-Civita symbol. \\
\hline
(149) & Effective equation of motion for $\kappa$ treated as a dynamic bilinear source. \\
\hline
(150) & Chain rule for operator mixing during RG flow of bilinear tensor contractions. \\
\hline
(151) & Trace projection of Yukawa–$\kappa$ mixing at one-loop level. \\
\hline
(152) & Divergence of a composite tensor operator; important in anomaly and flow analysis. \\
\hline
(153) & Coupling of field strength with quadratic flavor tensor structure. \\
\hline
(154) & CP-odd component arising from dual contractions of tensor operators. \\
\hline
(155) & Gauge–flavor tensor interaction under derivative mixing. \\
\hline
(156) & Gravitational curvature correction to $\kappa$ operator evolution. \\
\hline
(157) & Dimension-6 operator RG flow in four-vector covariant form. \\
\hline
(158) & Explicit form of higher-dimensional seesaw-induced operators in effective Lagrangian. \\
\hline
(159) & RG evolution of kinetic terms with curvature operator commutation. \\
\hline
(160) & Generalized divergence of $\kappa$ trace under geometric evolution. \\
\hline
(161) & Divergence of cubic tensor traces—relevant for flavor-violating predictions. \\
\hline
(162) & Tensor index contraction rules leading to kinetic norm conservation. \\
\hline
(163) & Box operator applied to $\kappa^2$ terms; leads to energy-momentum dependent effects. \\
\hline
(164) & Field strength-like structure constructed from non-Abelian $\kappa$ fields; hints at possible self-interaction. \\
\hline
\end{tabular}
\caption{Interpretation of high-dimensional and tensorial operator equations used in the analysis of radiative corrections and effective field evolution in the Type II seesaw framework.}
\label{tab:interpretation}
\end{table}

\section*{Appendix E: Feynman Diagrams of Operator Structures}

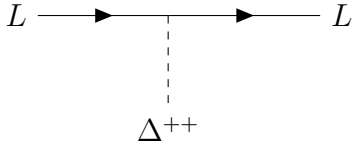
\begin{figure}[h!]
\centering
\begin{tikzpicture}
\begin{feynman}
\vertex (a) {\(L\)};
\vertex[right=2cm of a] (b);
\vertex[right=2cm of b] (c) {\(L\)};
\vertex[below=1.2cm of b] (d) {\(\Delta^{++}\)};
\diagram* {
  (a) -- [fermion] (b) -- [fermion] (c),
  (b) -- [scalar] (d),
};
\end{feynman}
\end{tikzpicture}
\caption{Dimension-5 neutrino mass operator via triplet scalar exchange in Type II seesaw: \(L L \to \Delta^{++} \to L L\).}
\label{fig:dim5}
\end{figure}

\begin{figure}[h!]
\centering
\begin{tikzpicture}
\begin{feynman}
\vertex (a1) {\(L\)};
\vertex[right=1.5cm of a1] (b1);
\vertex[right=1.5cm of b1] (c1) {\(H\)};
\vertex[below=1.5cm of b1] (d1) {\(\Delta\)};
\diagram* {
  (a1) -- [fermion] (b1) -- [fermion] (c1),
  (b1) -- [scalar] (d1),
};
\end{feynman}
\end{tikzpicture}
\caption{Dimension-6 operator mixing: radiative correction to \(LH\) vertex through \(\Delta\) insertion.}
\label{fig:dim6}
\end{figure}
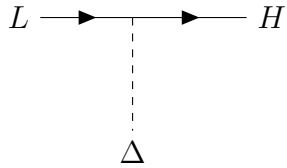

\end{document}